\documentclass[conference]{IEEEtran}
\IEEEoverridecommandlockouts
\usepackage{cite}
\usepackage{amsmath,amssymb,amsfonts}
\usepackage{algorithmic}
\usepackage{graphicx}
\usepackage{textcomp}
\usepackage{xcolor}
\def\BibTeX{{\rm B\kern-.05em{\sc i\kern-.025em b}\kern-.08em
    T\kern-.1667em\lower.7ex\hbox{E}\kern-.125emX}}

\PassOptionsToPackage{hyphens}{url}\usepackage[pdftex, colorlinks=true, hyperfootnotes=false, hyperindex=true, plainpages=false, pagebackref=false, pdfpagelabels=true, pdfstartview=FitH, linkcolor=blue, citecolor=blue, urlcolor=blue, bookmarks=false]{hyperref}

\usepackage{ifthen}

\newcommand{\note}[2]{}
\newboolean{showcomments}
\setboolean{showcomments}{false} 
\ifthenelse{\boolean{showcomments}}
  {\renewcommand{\note}[2]{
	\fbox{\bfseries\sffamily\scriptsize#1}
    {\sf\small$\Rightarrow$\textit{#2}$\Leftarrow$}
   }
  }
  {\renewcommand{\note}[2]{}
  }

\newcommand\iw[1]{\note{Ingo}{#1}}
\newcommand\ho[1]{\note{Heinrich}{#1}}
\newcommand\hl[1]{\note{Hendrik}{#1}}

\begin{document}

\title{FhGenie: A Custom, Confidentiality-preserving\\
Chat AI for Corporate and Scientific Use}

\author{	
\IEEEauthorblockN{
Ingo Weber\IEEEauthorrefmark{1}\IEEEauthorrefmark{2},
Hendrik Linka\IEEEauthorrefmark{1},
Daniel Mertens\IEEEauthorrefmark{1},
Tamara Muryshkin\IEEEauthorrefmark{1},
Heinrich Opgenoorth\IEEEauthorrefmark{1},
Stefan Langer\IEEEauthorrefmark{1}
}

\IEEEauthorblockA{
	\IEEEauthorrefmark{1} Fraunhofer Society, Headquarters, 80686 Munich, Germany
}
\IEEEauthorblockA{\IEEEauthorrefmark{2} School of CIT, Technical University of Munich, 
80333 Munich, Germany
}
\IEEEauthorblockA{
	Email: $\langle$firstname$\rangle$.$\langle$lastname$\rangle$@zv.fraunhofer.de
}
\IEEEauthorblockA{\IEEEauthorrefmark{2} ORCID: 0000-0002-4833-5921
}
}

\maketitle

\begin{abstract}
Since OpenAI's release of ChatGPT, generative AI has received significant attention across various domains. 
These AI-based chat systems have the potential to enhance the productivity of knowledge workers in diverse tasks. However, the use of free public services poses a risk of data leakage, as service providers may exploit user input for additional training and optimization without clear boundaries. Even subscription-based alternatives sometimes lack transparency in handling user data.

To address these concerns and enable Fraunhofer staff to leverage this technology while ensuring confidentiality, we have designed and developed a customized chat AI called \textit{FhGenie} (genie being a reference to a helpful spirit).
Within few days of its release, thousands of Fraunhofer employees started using this service.
As pioneers in implementing such a system, many other organizations have followed suit.

Our solution builds upon commercial large language models (LLMs), which we have carefully integrated into our system to meet our specific requirements and compliance constraints, including confidentiality and GDPR. 
In this paper, we share detailed insights into the architectural considerations, design, implementation, and subsequent updates of FhGenie. Additionally, we discuss challenges, observations, and the core lessons learned from its productive usage.
\end{abstract}


\section{Introduction}














Since OpenAI's release of ChatGPT, generative AI has gained significant attention across various domains. 
These AI-based chat systems have the potential to enhance the productivity of knowledge workers in diverse tasks. \iw{oneusefulthing - do not restrict user tasks; top-level management won't know best.}

ChatGPT, Google's Bard / Gemini, Anthropic's Claude, and similar tools are free to use in their basic version. However, the use of free public services poses a risk of data leakage, as service providers may exploit user input for additional training and optimization without clear guidelines. Even subscription-based alternatives sometimes lack transparency in handling user data. \hl{Ref needed?}

Fraunhofer is a research agency comprised of about 70 institutes, working in different areas of contract research with industry partners and projects. Scientists and other employees at Fraunhofer in their practical work have many use cases for generative AI. In addition, Fraunhofer has an overall need to be at the forefront of AI research and development, to be able to offer consulting services in that area to industry partners. 

\begin{figure}[tb]
    \centering
    \includegraphics[width=\columnwidth]{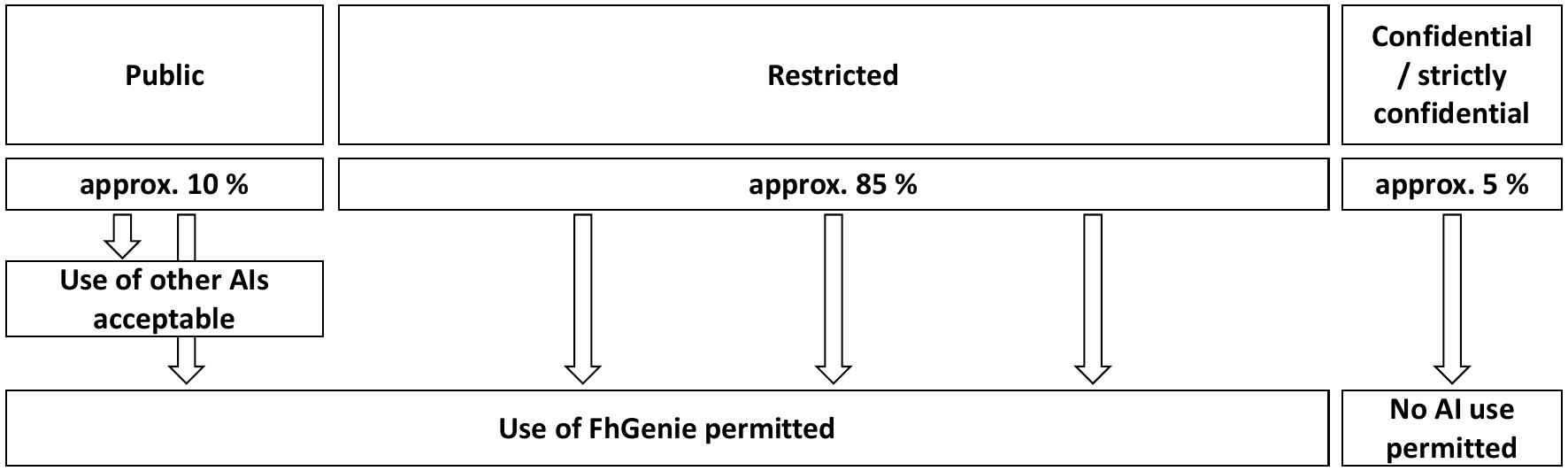}
    \caption{Information classification and use of AI}
    \label{fig:InfoClassification}
\end{figure}

\ho{At Fraunhofer, we use a system for information classification based on four categories, as shown ...}
At Fraunhofer, we have four classes in our information classification, as shown in \autoref{fig:InfoClassification}: \textit{public}, \textit{restricted}, \textit{confidential}, and \textit{strictly confidential}. After an initial assessment of AI chat tools and their terms and conditions, we came to the 
\ho{decision, that for the time being, only data classified as... ("momentary classification" gefällt mir nicht so :-)}
momentary classification, that only data classified as \textit{public} may be used in conjunction with these services.\footnote{Newer offerings for enterprises offer more favorable terms and conditions, but this was the situation when we first embarked on the FhGenie journey.} Even though Fraunhofer is an in-part publicly funded research organization, estimates of the percentage of data classified as \textit{public} are at around 10\% -- for companies, this percentage is likely lower. 
Even in scientific research as a domain of work, risks of data leakage can be high~\cite{DFG-2021}.

\ho{There are two issues with AIs that should only be used on \textit{public} data: (i) ...; (ii) most employees comply with the regulations and only use
 the AI for \textit{public} data -- but this is severely restring the usefulness ...}
The issue with AIs that should be only used on \textit{public} data is as two-fold: (i) some employees might use such AIs for \textit{restricted} contents, leaking sensitive data and possibly being non-compliant with regulations like GDPR and the Data Act; (ii) most employees do not do that, thereby restricting the usefulness of the new tools and the level of familiarity and testing they do with generative AI.
Accordingly, we were highly motivated to create an alternative in which data and matters classified as \textit{restricted} could also be used. 
With FhGenie, we achieved that goal, rendering its use permissible for 95\% of matters as opposed to 10\%.

FhGenie is a combination of FhG, the abbreviation of Fraunhofer Gesellschaft, and genie, the useful spirit. The initial release of FhGenie was made public in a press release~\cite{FhGenie:PR:2023}, but architectural and technical information was not the focus of this release.
We believe the insights and lessons learned to be of general interest to many organizations, hence in this paper we share our architectural considerations and insights gained from designing, developing, and operating a generative AI tool.
Our chat AI utilizes the Microsoft offering ``Azure OpenAI Service'', giving us access to state-of-the-art AI models.
FhGenie is being extended with Fraunhofer-specific data to cater for additional usage scenarios, while at the same time being isolated from OpenAI’s and Microsoft’s public cloud and services. With this design, we achieve a high level of confidentiality and compliance in accordance with the European regulations. 

The remainder of the paper is organized as follows. In the next section, we discuss the genesis of FhGenie. 
Subsequently we present the requirements and the resulting design in Sections~\ref{sec:reqs} and \ref{sec:design}.
Central aspects of the development, deployment, operation, use, and user feedback are discussed in \autoref{sec:DevOpsUsage}, before \autoref{sec:future} covers ongoing and future developments.
\ho{We conclude the paper with a summary and brief discussion in \autoref{sec:summary}.}
We end the paper with a summary and short discussion in \autoref{sec:summary}.

\section{Genesis and Evolution of FhGenie}
\label{sec:genesis}
The release of ChatGPT with the LLM GPT-3.5 in November 2022 and of GPT-4 in March 2023 demonstrated to the world the state of the art, clearly outperforming then-current open-source / open-parameter models and competitors~\cite{2024-LLMs-for-Science,phung2023generative,bai2023longbench}. 
Accordingly, to build a sufficiently competitive chat AI, access to the models was required.

GPT-3.5-Turbo models became available through Azure in May 2023\footnote{\url{https://learn.microsoft.com/en-us/azure/ai-services/openai/whats-new}, accessed 2024-02-17}. When learning of this development, we decided to implement FhGenie on this basis, with a focus on speed. In parallel to designing and implementing FhGenie, we sought consensus with the workers' council -- due to German labor laws, this council may have had the right to veto the introduction of the new tool, a risk we mitigated by finding consensus. 
The design of FhGenie is influenced by considerations to minimize the possibly negative effects of such tools on workers, as we will detail below. By preempting acceptable solutions for likely workers' concerns, we were able to achieve consensus fast. The implementation of the first version of FhGenie also progressed fast, so that we were able to go live with the tool at the end of June.

This first version was limited in a number of ways, and through subsequent iterations (involving discussions with the workers' council representatives, design, development, and deployment) we made a number of improvements and are working on others.
\ho{Vorschlag:
This first version was limited in a number of ways, and through subsequent iterations (involving discussions with the workers' council representatives and the teams responsible for design, development, and deployment) we made a number of improvements and are working on others.
}
Such improvements include the extension to other models (GPT-4, GPT-4-Turbo, DALL-E), retrieval-augmented generation (RAG) based on Fraunhofer-specific data, and API access for scientists.
In the following, we take the viewpoint of today's design (unless explicitly stated otherwise) since we believe this to be of the highest interest for readers.

\section{Requirements}
\label{sec:reqs}
From conversations with (prospective) users and other stakeholders, and on the basis of our IT landscape, we compiled the following list of requirements:
\begin{itemize}
    \item Integration with existing user management: user authentication and authorization should be based on our existing Single-Sign-On (SSO) system; institute administrators have to explicitly authorize which user groups / users are authorized.
    \item State-of-the-art AI models: the tool should make use of existing models.
    \item Confidentiality: user prompts and responses must be kept at an appropriate level of confidentiality for \textit{restricted} data.
    \item Secure sandbox: no interaction should allow escaping the confines of the application; leaking of input or output (even indirect via model training/fine-tuning) from a user session must be prevented, unless content is marked to indicate otherwise.
    \item Compliance: to ensure compliance of data handling (EU GDPR, EU Data Act), no data should leave the EU.
    \item Ease of use, responsive user interface (UI): the UI should be as intuitive as commercial tools, and support screen sizes ranging from smart phones to large desktop monitors; language support for German and English.
    \item API access: some scientists desired API access for use in their research project.
    \item Priority of support for selected modalities: the focus was on natural language prompts and responses, with a secondary goal of supporting program code input and output; image input and output are interesting, but less important; video and audio are out of scope for the time being.
    \item Responsible AI, privacy: malicious prompts should be detected and logged; user input is private and individual usage and content are not tracked by managers or other parts of the organization; AI may not be used to judge people (job applications, promotion, bonus, etc.).
    \item Cost-effective, with acceptable latency and bandwidth: latency should be comparable to commercial solutions, i.e., most natural language prompts should be answered within few seconds, but processing complex prompts and long inputs may take longer (10-20 seconds);
    bandwidth should be sufficient for regular use by approx.\ 30,000 users (all employees at Fraunhofer); cost should be acceptable within the budget set (exact number is confidential, but it is lower than commercial subscriptions for all users), and cost must be monitored.
\end{itemize}

\section{Architecture Design}
\label{sec:design}

\begin{figure*}[tb]
    \centering
    \includegraphics[width=0.85\textwidth]{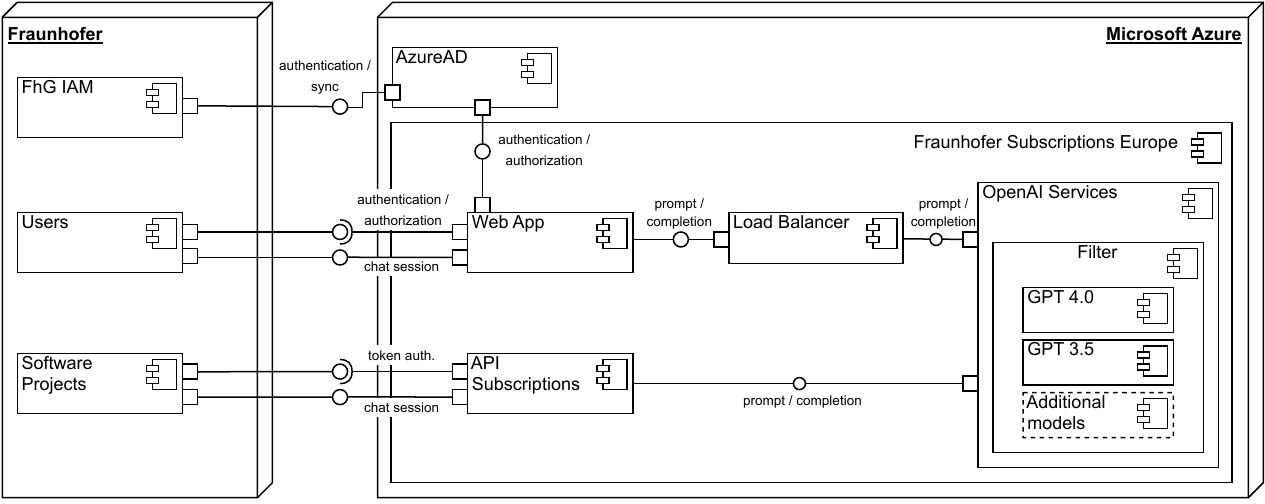}
    \caption{FhGenie architecture overview (UML component diagram)}
    \label{fig:arch_overview}
\end{figure*}

Based on the above requirements, we designed the architecture of FhGenie depicted in \autoref{fig:arch_overview}.
The majority of components is hosted within Microsoft Azure, specifically the Fraunhofer subscriptions in Europe, which offer isolation and security properties that meet the respective requirements, including compliance.
This also applies to the AI models within the OpenAI Services component on the right-hand side.
In particular, neither Microsoft nor OpenAI use the information from Fraunhofer users for model fine-tuning or model training, thus preventing the indirect leakage of sensitive information that could occur otherwise.

Fraunhofer users (left-hand side of \autoref{fig:arch_overview}) interact with the web application hosted on Azure, which realizes authorization and authentication through the Azure active directory (AzureAD, the product ``Entra ID'').
The AD in turn is connected to Fraunhofer's identity and access management (FhG IAM), through which user and rights information is collected from various sources (e.g., information about employees joining and leaving from HR systems, specific rights from administrators at the institutes). The relevant information is synchronized with the AD, which is a logically centralized directory for all Azure subscriptions and other services not shown in the figure.
Authentication of users is always against the FhG IAM systems, while the AD takes care of authorisation.

User inputs to the web application are turned into prompts to the model, with some level of prompt engineering in the backend (described in more detail in the next section).
In response to a prompt, the model provides a completion that is turned into output to the user. 
The conversion in both directions includes format changes, e.g., program code is formatted differently than natural language text.
The UI was designed to work well for a range of screen sizes. 

\begin{figure*}[tb]
    \centering
    \includegraphics[width=0.8\textwidth]{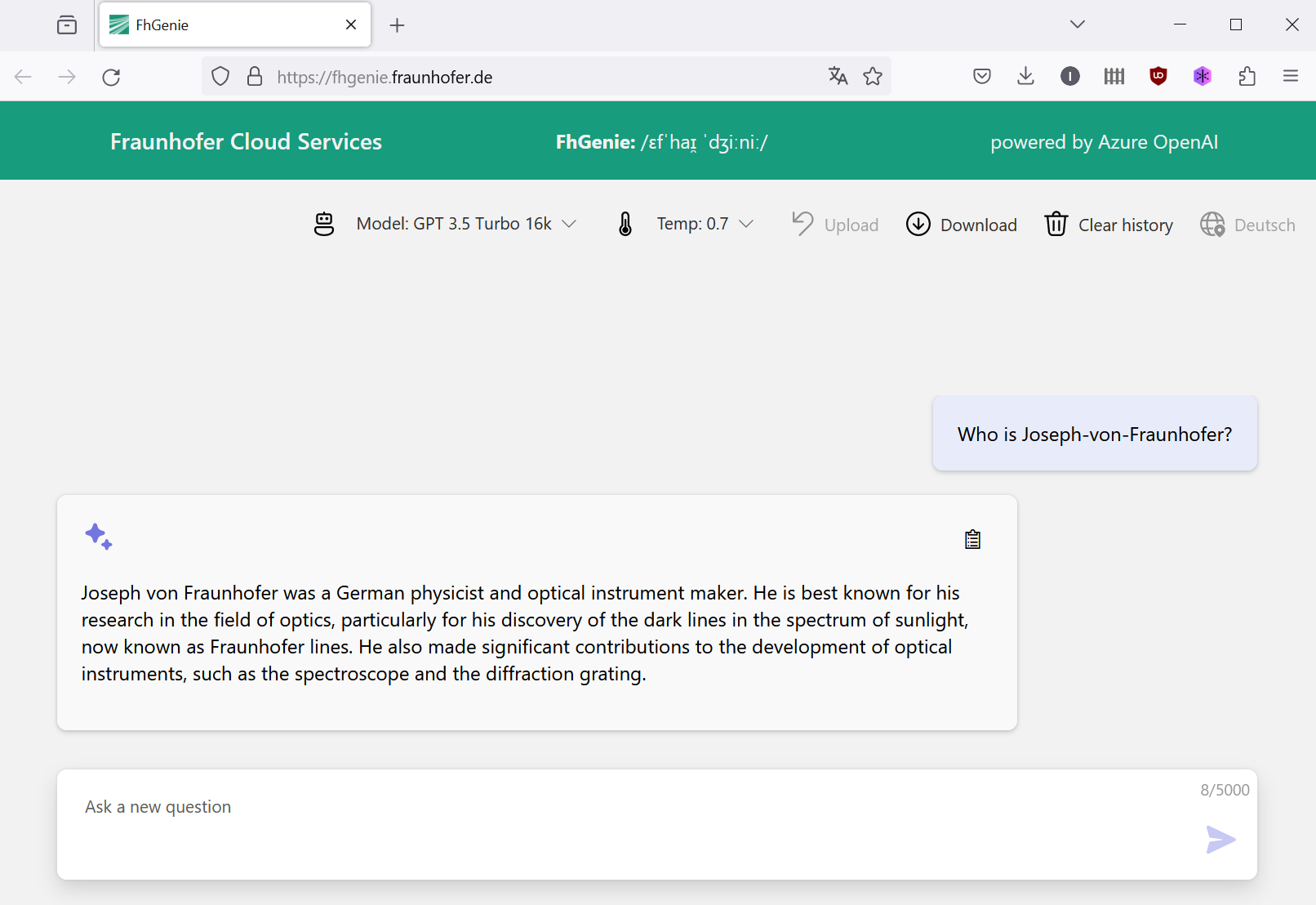}
    \caption{Screenshot of FhGenie}
    \label{fig:screenshot}
\end{figure*}

A screenshot of the UI is shown in \autoref{fig:screenshot}.
Users can select the AI model to use and the language (German or English). They can further opt to download the state of a conversation and store it locally; having done so earlier, they can upload that conversation at a later time and continue it. While this method is less comfortable than storing the conversation on the server, it meets the privacy requirement. Users can also clear the current conversation.
Finally, the users can also control the ``temperature'', a model parameter between 0 and 1 that makes the output more or less deterministic. Lower values correspond to more deterministic output, which is better for factual contexts, but may be too limited for conversing about ideas. Higher values increase the likelihood of hallucination, or rather: confabulation~\cite{Smith:2023:Confabulation}.
Offering such options is a deliberate choice for our context, where many of the users are research scientists.
A different choice might suit other professional environments better.
We limit the user input to 5000 characters, which is made transparent to the users in the input field.

While the chosen design was acceptable in terms of latency, bandwidth, and cost with GPT-3.5, this changed with the switch to GPT-4. Unfortunately the number of tokens per minute that we can access through our use of Azure is a limiting factor in the EU.
To counter this effect, we introduced load balancing between different AI model API endpoints across the EU.
With this mechanism, we could achieve sufficient bandwidth and low-enough latency at acceptable cost.

In parallel to the users interacting with the web application, software projects from Fraunhofer scientists may also interact with an API that we provide. Such interaction is typically bound to a research project, not an individual user -- hence, authentication is achieved by means of security tokens. Tokens are issued to projects individually and transmitted via encrypted email (not shown in the figure).
The number of API clients is considerably smaller at this time, such that spreading them out over multiple instances is sufficient at this time; in the future, we might adapt the load balancer for API clients and introduce it there as well.

Generative AI can certainly be used for desirable purposes, such as productivity gains and support for menial tasks~\cite{peng2023impact,le2023chatgpt}, but it can also be used in harmful ways~\cite{walsh2018-2062-book}\footnote{\url{https://www.nytimes.com/2024/01/26/arts/music/taylor-swift-ai-fake-images.html}, accessed 2024-02-18}\footnote{\url{https://securityconference.org/aielectionsaccord/} and \url{https://news.microsoft.com/2024/02/16/technology-industry-to-combat-deceptive-use-of-ai-in-2024-elections/}, both accessed 2024-02-18}.
Hence, the use of approaches from responsible AI (RAI, \cite{RAI-book}) is important.
Azure OpenAI Services offers a base level of RAI, including filters and an abuse log, where seemingly harmful interactions are stored securely. It is possible to customize this system, but that has implications on the split of tasks and responsibilities between Microsoft and the customer, i.e., us. The exact implications of doing so need to be understood and considered before any customization. Based on our assessment in May 2023, we chose to rely on the default RAI system, which is adequate for our internal service. We will revisit this architectural choice at a later time.
This system does not prevent prompts that aim to have people judged by the AI by technical means, and we did not find mature technical solutions to meet this requirement. Accordingly, we applied non-technical means (note on the landing page of FhGenie, guidelines, email communication) to inform users about responsible use of AI.
This includes information about the use in the context of scientific work.

\section{Development, Operations, and Usage}
\label{sec:DevOpsUsage}

The code base of FhGenie is written in NodeJS and Python, using the Flask web framework, and is split into two parts: the frontend and the backend.
The backend is instantiated twice: once for the frontend and once for API access of software projects.
It encapsulates the AI models, which are spread over two or more regions depending on resources available from Azure within the EU. The backend for the UI also encapsulates the load balancer and prompt engineering.
The prompt engineering operates as follows: text from a given user question is enriched with context information and instructions before it is sent as a prompt to the AI model.
We experimented with a number of different types of context / instructions, and found that overly specific instructions had adverse effects for our general-purpose AI chat. Therefore, our current system prompt contains about eight instructions. More context-specific AI systems will likely require more prompt engineering and result in longer system prompts.
The load balancer is based on a publicly available code sample\footnote{\url{https://github.com/Azure-Samples/openai-apim-lb}, accessed 2024-02-18}, which we adapted to our context. If an AI model's limit regarding token or request throughput is reached, the load balancer detects this and redirects traffic to other instances.
Users of the FhGenie API share instances. If this results in adverse effects, we enter a conversation with the developers and try to find better solutions.

Our frontend implementation is also based on a publicly available code sample\footnote{\url{https://github.com/Azure-Samples/azure-search-openai-demo}, specifically the version from 2023-05-17: \url{https://github.com/Azure-Samples/azure-search-openai-demo/tree/01a88b2f8aeaaa19c7613b4c373605bdda21fed2}, both links accessed 2024-02-18}.
The version we used was not robust enough for productive use in our context, such that e.g.\ poor internet connection led to users losing all progress in a conversation. We also added most of the UI features mentioned in the previous section. Furthermore, the first version was vulnerable to malicious code injections, e.g., through prompts asking for Javascript code responses, which were executed on the client.

When deploying the system, we make use of infrastructure-as-code (IaC), specifically using Azure Bicep.
Not all steps are automated though, e.g., the AI model instances are still created manually for the time being.
For runtime monitoring and management, we make use of Azure mechanisms for token and cost budgets.
During operation, we monitor cost, usage (number of requests and tokens), and other aspects like health of the machines.
One metric we are monitoring is the number of blocked requests -- which is typically at 0 in a given week.
For Fraunhofer scientists whose research requires testing the limits and harmful uses of generative AI, we opened up other channels to access AI models.
We also respond to questions and issues raised by users, including requests for solutions for specific contexts.
In addition, we maintain an intranet page and a Q\&A section. 

In the week following the initial release of FhGenie, over 6000 users tested the service. After two months, we had approx.\ 7000 active users. Not all staff have access, which is controlled in a decentralized way by administrators of the research institutes. At the time of writing, over 25,000 of the 30,000 staff of Fraunhofer are authorized to use FhGenie.
We observe approx.\ 10,000 requests per day.


User feedback was very positive regarding the availability of FhGenie, particularly with regards to the speed of the initial release.
A common theme in user feedback was that, whenever new features were added to commercial tools like ChatGPT, FhGenie users would ask whether we would or could add such features.
This included new or extended AI models, plugins, and custom GPTs.
Other frequent user requests included API access and increasing the prompt size.
We were able to satisfy many of these desires, but we decided against implementing some. The latter category includes custom GPTs and plugins.
In the former category, we went from GPT-3.5-Turbo with a 4k token window and no API access to
GPT-3.5-Turbo with a 16k window, GPT-4, and API access.
As for non-technical requests, users frequently asked for help how to make use of FhGenie, both in general and in their specific work area.
While we could answer some questions, we address the demand at large by building a community of practice in an internal exchange medium. This is an ongoing effort, for which we are currently evaluating different options regarding its organization and tooling.

\section{Ongoing and Future Work}
\label{sec:future}

FhGenie is subject to ongoing development -- in the following, we list a number of avenues that we are currently exploring or are going to explore in the future. The directions are sorted in a roadmap fashion, i.e., from ongoing to more distant future directions.

\begin{itemize}
    \item Feeding \textit{Fraunhofer-specific data} into FhGenie: in order to offer more useful, context-specific replies to our scientists and other knowledge workers than currently, we need to augment our AI model instances with organization-specific data. Of the available methods to achieve this, we decided to pursue an approach called ``retrieval augmented generation'' (RAG, \cite{Lewis:2020:RAG}). 
    To implement RAG, 
    we augment our architecture with a semantic index,
    which is a database used to generate prompt context (i.e., it enriches the prompts with knowledge from documents) for specific queries to the LLM. 
    This feature is under active development, and is expected to be ready for productive use in the coming weeks or months.
    In terms of data selection, we plan to add content from out intranet pages as well as from public web pages, the latter of which containing current information of research areas and expertise from the different Fraunhofer institutes. 
    Many other data sources might also be of interest, such as databases of Fraunhofer publications, patents, data sets, etc., and we will explore the value of adding those into our RAG extension in due time.

    \item \textit{Additional generative features or modalities} from Azure:
    Microsoft is constantly adding more features like image generation to their company-specific OpenAI offerings, and of course we consider for each of those whether and when to make these available to our users. 
    As discussed in the requirements section, modalities like image generation are interesting for us. 
    We consider the new features and modalities, and decide whether to make these available to our users. 
    Customization might also be valuable here, e.g., an image generator with a large body of corporate design context might allow unlocking new areas of productivity gains. 
    The main problems currently are resource limitations on the Azure services, which make e.g.\ image generation impractical for our user base.
    For additional modalities and features that we consider, trade-off analyses considering value, cost, and other factors need to be made.
    
    \item Using \textit{prompt compression} to reduce cost and increase performance: Microsoft faces issues serving the demand of their large customer base, which results in bandwidth limits in terms of number of tokens and requests per minute. 
    The way we access the AI models through Azure for FhGenie is cost-efficient, but the limits are lower than we would wish.
    This is especially relevant for prompts with a large amount of context data. To mitigate this problems and potentially reduce cost (based on tokens used), we are investigating prompt compression with LLMLingua\footnote{\url{https://www.llmlingua.com/}, accessed 2024-02-18}\cite{jiang-etal-2023-llmlingua,jiang-etal-2023-longllmlingua}. The core idea is to use a smaller LLM to pre-process the input, resulting in a smaller and more efficient prompt which then is processed by the larger and more costly model like GPT-4. The smaller LLMLingua model could be run on local GPUs, which can be more cost-efficient than using cloud offerings.
    This is a complex and speculative direction, and as yet it is unclear if fit for productive use. If successful, it could alleviate pain and cost, and these advantages might apply to any use of LLMs.
    Alternative paths include switching to more expensive service options at Azure and spreading traffic over different models, including the options discussed next.

    \item Using \textit{other AI models}: both the industry and research labs are releasing new models, be it as open-source / open-parameter models or closed models behind APIs\footnote{For instance, Google Gemini API: \url{https://ai.google.dev/models}, accessed 2024-02-18}. Notably, some advanced open-parameter models are created and released by companies such as the LLama models by Meta\footnote{\url{https://llama.meta.com/}, accessed 2024-02-18}. 
    While alternative models are in principle subject to the same considerations and trade-offs as Azure features and modalities, utilizing them in FhGenie would require considerable changes to the architecture.
    For open-parameter models, questions around RAI, RAG and underlying hardware have to be revisited; some are available from cloud providers, while others would require procuring or renting GPU capacities along with the operation.
    As for RAI and RAG, we are investigating model-independent mechanisms to provide guardrails and a semantic index. If we succeed with this endeavor and the quality attribute trade-offs are positive, we might achieve very high maintainability: switching the AI model in the backend would become relatively easy, while customizations in RAG and RAI would be independent of it.
\end{itemize}


\section{Summary and Discussion}
\label{sec:summary}

In this paper, we shared our insights into the motivation, architecture goals and design, development and operation of an LLM-based chat AI, FhGenie. 
The tool relies on state-of-the-art LLMs, specifically GPT-3.5 and GPT-4, provisioned through Microsoft Azure.
To Fraunhofer staff, it offers a UI and an API, which are integrated with our SSO tooling.
Our design meets quality requirements, such as bandwidth, latency, cost, and responsible AI.
FhGenie is used by thousands of Fraunhofer staff to make use of generative AI technology.  

FhGenie is one piece in a larger puzzle of AI technologies being evaluated and introduced at Fraunhofer.
The goal is not to compete with the market leaders, but to offer employees a good, compliant and secure alternative to commercial solutions. 
Generative AI with current capabilities is a new technology, and besides direct usefulness. Another goal is to enable staff to experiment and understand this technologies, what it can do and where its use does not yield benefits.
These goal has been achieved.
\ho{What is the intention here, "These goals have" or "this goal has" been achieved?}
\iw{oneusefulthing}

Ongoing and future developments include additional modalities, such as image generation, RAG on internal documents, advanced optimization techniques like prompt compression, and exploring architectural changes to enable switching of underlying AI models.

\section*{Acknowledgments}
We thank FhGenie for its feedback on earlier versions of the text. Note that none of the text in this paper was generated by an AI, though.
We further thank Frank Strathoff and Axel M{\"u}ller-Groeling (Fraunhofer) as well as Microsoft for their support during the design, implementation, and operation of the service.

\bibliographystyle{plain} 
\bibliography{refs} 

\end{document}